\documentclass[a4paper]{jpconf}
\usepackage{graphicx}
\begin{document}
\title{$\psi$(2S) Production at the LHC}

\author{Xiaojian Du and Ralf Rapp}
\address{Cyclotron Institute and Department of Physics \& Astronomy, \\
Texas A\&M University, College Station, TX 77843-3366, USA}
\ead{xjdu@physics.tamu.edu, rapp@comp.tamu.edu}

\begin{abstract}
We calculate the production of $\psi(2S)$ and the pertinent double ratio of its nuclear
modification factor ($R_{\rm AA}$) over that of the $J/\psi$ in Pb-Pb collisions at
the LHC. Based on a transport model with temperature dependent
reaction rates, a sequential regeneration pattern emerges: the larger $\psi(2S)$ width,
relative to the $J/\psi$, around and below the critical temperature, implies that most of
the $\psi(2S)$ states are regenerated later in the evolution of the fireball. This has
noticeable consequences for the transverse-momentum ($p_T$) spectra of the regenerated
charmonia.
While the total yield of $\psi(2S)$ meson remains smaller than those of $J/\psi$'s, their
harder $p_T$ spectra can produce a double ratio above unity for a $p_T>3$\,GeV cut, as
applied by the CMS collaboration. A significant uncertainty in our calculations is
associated with the values of the temperature where most of the $\psi(2S)$ regeneration
occurs, {\it i.e.}, the quantitative temperature dependence of its inelastic width.
\end{abstract}

\section{Introduction}
Intense experimental efforts are ongoing to measure the production systematics of heavy quarkonia
in heavy-ion collisions (HICs), to establish a robust benchmark for studying their modifications
in hot and dense QCD matter~\cite{Kluberg:2009wc,BraunMunzinger:2009ih,Rapp:2008tf}. The sequence of bound states in the vacuum spectrum of heavy quarkonia
provides a unique probe of how the basic QCD force evolves in the medium. In the charmonium sector,
where extensive measurements of $J/\psi$ production have revealed important information about its
in-medium kinetics, the $\psi(2S)$ state is now becoming a rather hot topic. A strong $\psi(2S)$ suppression
relative to the $J/\psi$ has been observed in both proton-nucleus (pA) collisions at RHIC and the
LHC~\cite{Adare:2013ezl,Adam:2016ohd}, and in nucleus-nucleus collisions~\cite{Alessandro:2006ju}
at the SPS, generally associated with final-state interactions together with a much smaller
binding energy of the $\psi(2S)$. It was therefore rather intriguing when the CMS collaboration
found an enhancement of the $\psi(2S)$/$J/\psi$ ratio in central Pb-Pb(2.76\,TeV) collisions,
relative to pp collisions~\cite{Khachatryan:2014bva}. More precisely, this enhancement was
found in the kinematic range of forward-rapidities (1.6$<$$|y|$$<$2.4) and transverse momenta
$p_T$$>$3\,GeV, while it turns out to be suppressed around mid-rapidity and $p_T$$>$6.5\,GeV.
In Ref.~\cite{Du:2015wha} we suggested a sequential regeneration of $J/\psi$
and $\psi(2S)$ states as a potential mechanism to understand this phenomenon within schematic
model scenarios. In the present paper we expand on this work by revisiting the results within
a more quantitative rate equation approach~\cite{Grandchamp:2003uw,Zhao:2010nk,Zhao:2011cv}
(Sec.~\ref{sec_rate}), discussing the double ratio and its main uncertainties in Pb-Pb(2.76\,TeV)
collisions (Sec.~\ref{sec_276}), and providing predictions at 5.02\,TeV (Sec.~\ref{sec_502}). We
conclude in Sec.~\ref{sec_concl}.

\section{Transport model for charmonia production}
\label{sec_rate}
Our approach for quarkonium kinetics in HICs utilizes a rate equation,
\begin{equation}
\frac{\mathrm{d} N_\Psi}{\mathrm{d}\tau}=-\Gamma_\Psi(T)\left[N_\Psi-N^{\rm eq}_\Psi(T)\right]
\ ,
\end{equation}
which evolves the charmonium yields, $N_\Psi$ ($\Psi$=$J/\psi,\psi(2S),\chi_c(1P)$)
through an expanding fireball. The pertinent transport coefficients are: (a) the
inelastic reaction rate $\Gamma_\Psi(T)$ in the quark-gluon plasma (QGP) and in hadronic
matter. ``Quasifree" dissociation is adopted as the dominant mechanism to break up $J/\psi$ and
$\psi(2S)$ states in the QGP where both have relatively small binding energies, $E_B \le T$, while
hadronic dissociation rate is calculated from a meson exchange model with $SU_f(4)$ flavor symmetry;
(b) the thermal equilibrium limit $N_\Psi^{\rm eq}(T)$ which controls the rate of regeneration.
It is evaluated from the statistical model in either partonic or hadronic basis and includes
non-equilibrium corrections due to a finite correlation volume and incomplete charm-quark
thermalization~\cite{Grandchamp:2003uw}.

To compute $p_T$ spectra, we decompose the solution of the rate equation into two parts,
one characterizing the primordial (direct) production and the other from regeneration of
$\Psi$ states.
The Boltzmann equation is used to calculate the $p_T$-spectra of the primordial part via
\begin{equation}
\frac{\partial f(\vec{x},\vec{p},t)}{\partial t}+\vec{v}\cdot\frac{\partial f}{\partial \vec{x}}
=-\Gamma_\Psi(\vec{p},T)f(\vec{x},\vec{p},t)
\end{equation}
where $f$ is the phase space distribution of charmonia and $\vec{v}=\vec{p}/{E_p}$ their velocity.
Escape effects are included by setting the rate to zero if a state exits the fireball boundary.
The $p_T$-spectra of the regeneration component are evaluated from a blastwave description assuming
thermal equilibrium,
\begin{equation}
\frac{\mathrm{d}N}{p_T\mathrm{d}p_T}\simeq m_T \int_0^R r\mathrm{d}r
K_1\left(\frac{m_T \mathrm{cosh}(\rho(r))}{T}\right)I_0\left(\frac{p_T\mathrm{sinh(\rho(r))}}{T}\right)
\end{equation}
where $m_T=\sqrt{p_T^2+m^2}$ is the transverse mass and $\rho(r)=\mathrm{tanh}^{-1}(v(r,t,b))$
the transverse-flow profile of the fireball. Later times in the evolution lead to a harder spectra,
as lower temperatures are overcompensated by the blue-shift due to larger flow.
The blastwave spectra are normalized to the regeneration yield obtained from the $p_T$-independent
rate equation. The pertinent $R_{\rm AA}$'s are computed in the usual way as a ratio of
AA and pp spectra, $R_{\rm AA}=(N^{\rm AA})/(N_{\rm coll}N^{\rm pp})$,
scaled by the binary collision number, $N_{\rm coll}$, obtained from the optical Glauber model.

The space-time evolution of the temperature is constructed from a simple ansatz for an iso-
and isentropically expanding fire cylinder volume, $V_{\rm FB}(t)$. Using conservation of
total entropy,
\begin{equation}
S_{\rm tot}=s(T)V_{\rm FB}(t) \ ,
\end{equation}
together with a suitable equation of state for the entropy density, $s(T)$ (for which we use a QGP
quasi-particle and hadron-resonance gas connected through a mixed phase at $T_c$=180\,MeV), the
time-dependent temperature is obtained for a given centrality by matching
$S_{\rm tot}$ to the final-state hadron multiplicities.

\section{Sequential regeneration in Pb-Pb(2.76\,TeV)}
\label{sec_276}
\begin{figure}[t]
\includegraphics[width=0.48\textwidth]{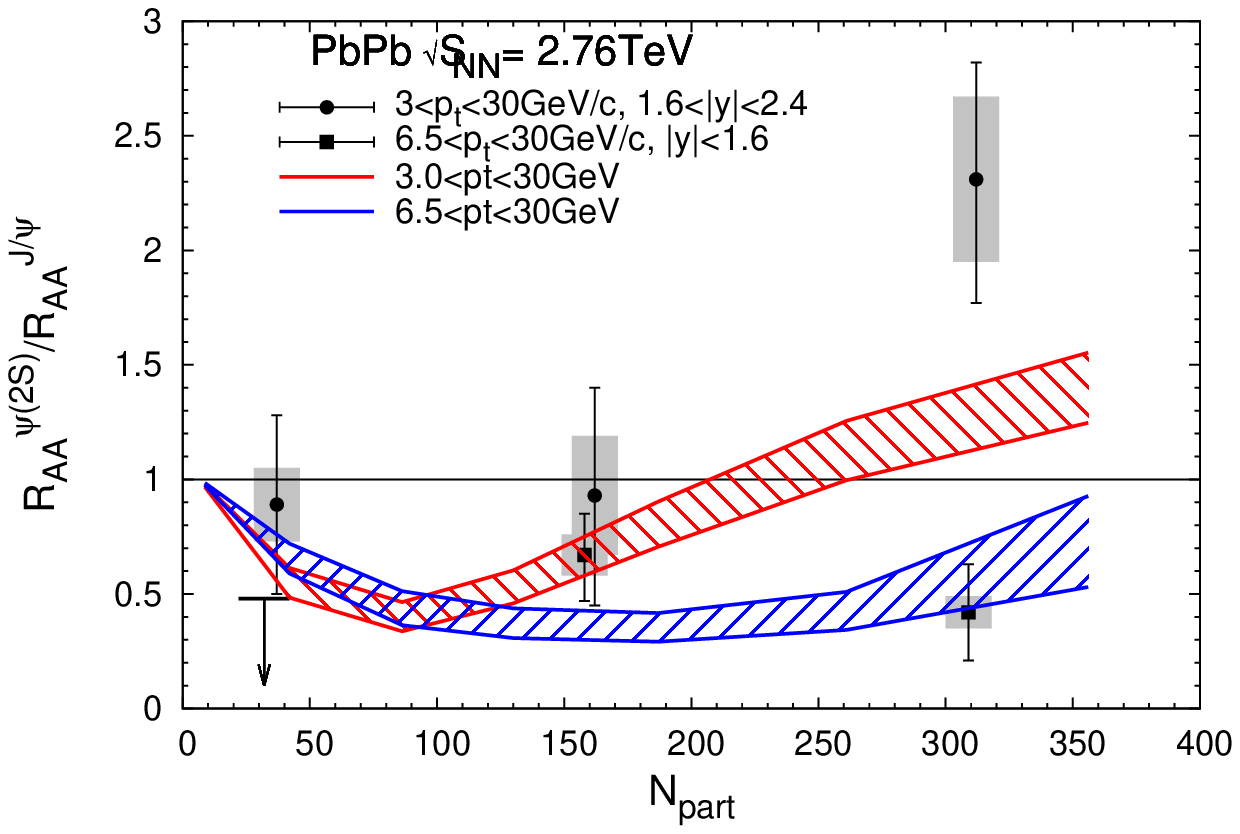}
\includegraphics[width=0.48\textwidth]{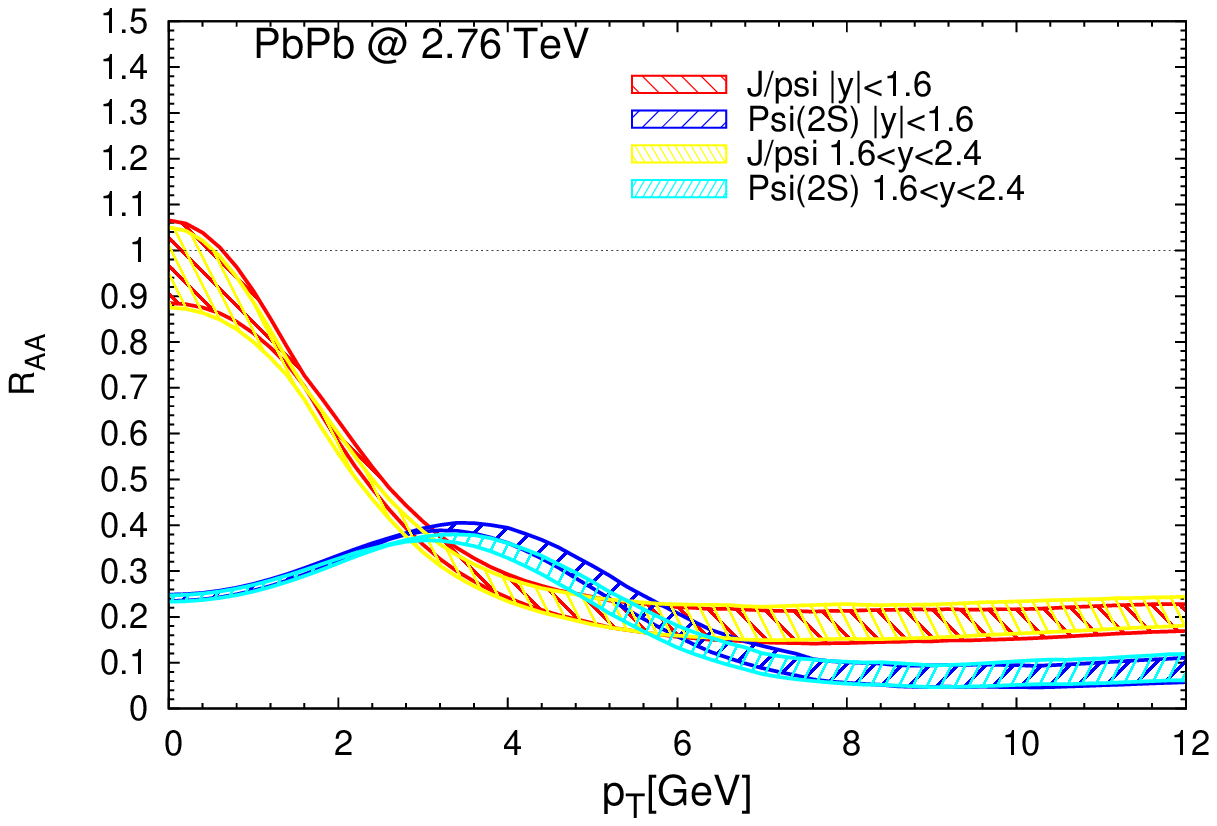}
\caption{Charmonium production in Pb-Pb collisions at $\sqrt{s_{NN}}$=2.76\,TeV within the kinetic
rate equation approach. Left panel: centrality dependence of the double ratio
$R_{\rm AA}(\psi(2S))/R_{\rm AA}(J/\psi)$ for $p_T$$>$6.5\,GeV and $|y|$$<$1.6 (blue band)
as well as $p_T$$>$3\,GeV and 1.6$<$$|y|$$<$2.4 (red band), compared to CMS
data~\cite{Khachatryan:2014bva}.
Right panel: $p_T$ dependence of the individual $J/\psi$ and $\psi(2S)$ $R_{\rm AA}$'s
for central collisions. A 10\,\% shadowing is assumed in the $p_T$-spectra according to EPS09 NLO~\cite{Eskola:2009uj}.}
\label{fig_276}
\end{figure}

Our results for the centrality dependence of the $R_{\rm AA}$ ``double ratio",
$R_{\rm AA}(\psi(2S))/R_{\rm AA}(J/\psi)$, as obtained from the kinetic rate equation
are displayed in Fig.~\ref{fig_276}. They confirm the results of our previously
published schematic-model study~\cite{Du:2015wha}; moderate quantitative deviations
arise from the more restrictive model approach which intimately couples the suppression
and regeneration yields. The basic trends of the CMS data in Pb-Pb(2.76\,TeV) collisions
for the two $p_T$ cuts are still reproduced (left panel of Fig.~\ref{fig_276}),
although the maximal enhancement in central collisions is not fully reproduced. Nevertheless,
the marked increase of the double ratio beyond one remains a key signature of the
{\it sequential regeneration} mechanism, whereby most of the $\psi(2S)$ are formed
later in the evolution, thus being blue-shifted to $p_T$ above 3\,GeV (red band). The
regenerated $J/\psi$ are mostly concentrated at momenta $p_T\le$3\,GeV, consistent with
ALICE data~\cite{Adam:2015isa}. On the other hand, for $p_T>$6.5\,GeV (blue band), the
regeneration component has essentially faded away (being exponentially suppressed
relative to the primordial power-law spectra), and the stronger suppression of the
primordial $\psi(2S)$ relative to the $J/\psi$ leads to a double ratio below one.
The explicit $p_T$ dependence of the $\psi(2S)$ and $J/\psi$ $R_{\rm AA}$s for
central Pb-Pb is depicted in the right panel of Fig.~\ref{fig_276}.

The largest contribution to the uncertainty bands in Fig.~\ref{fig_276}
is due to the choice of the average temperature, $\bar{T}_{\rm reg}$, at which the
blast-wave expression is evaluated, representing the window where most of the
regeneration occurs. From the time dependence of the regeneration yields we estimate
$\bar{T}_{\rm reg}$$\simeq$160-165\,MeV for the $\psi(2S)$ and $\bar{T}_{\rm reg}$$\simeq$180-200\,MeV for the $J/\psi$~\cite{Du:2015wha} which can describe the ALICE data for $R_{\rm AA}^{J/\psi}(p_T)$ at 2.76\,TeV~\cite{Abelev:2013ila}.
We also varied the initial spectra by inclusion of a moderate Cronin effect with
a broadening parameter of up to $a_{gN}$=0.2\,GeV$^2$/fm.

\section{Predictions for 5.02\,TeV}
\label{sec_502}
\begin{figure}[!t]
\includegraphics[width=0.48\textwidth]{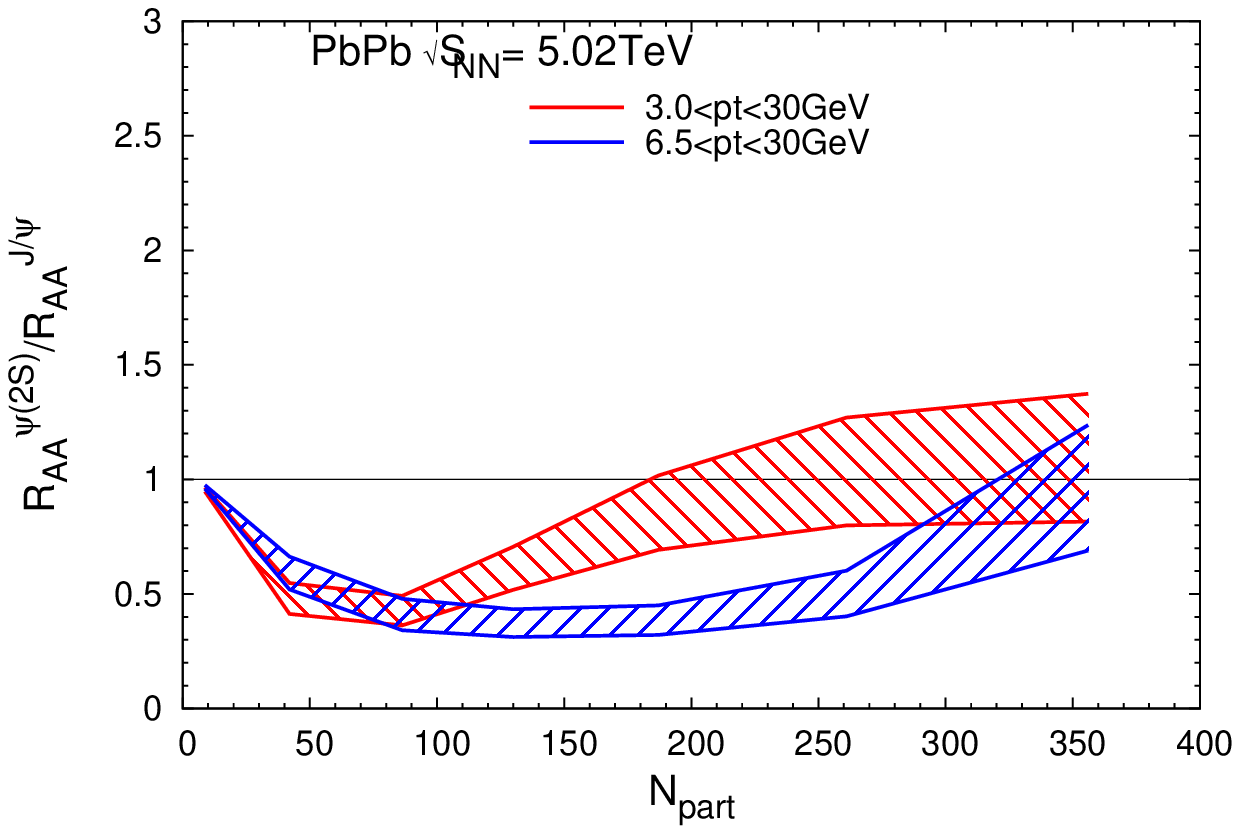}
\includegraphics[width=0.48\textwidth]{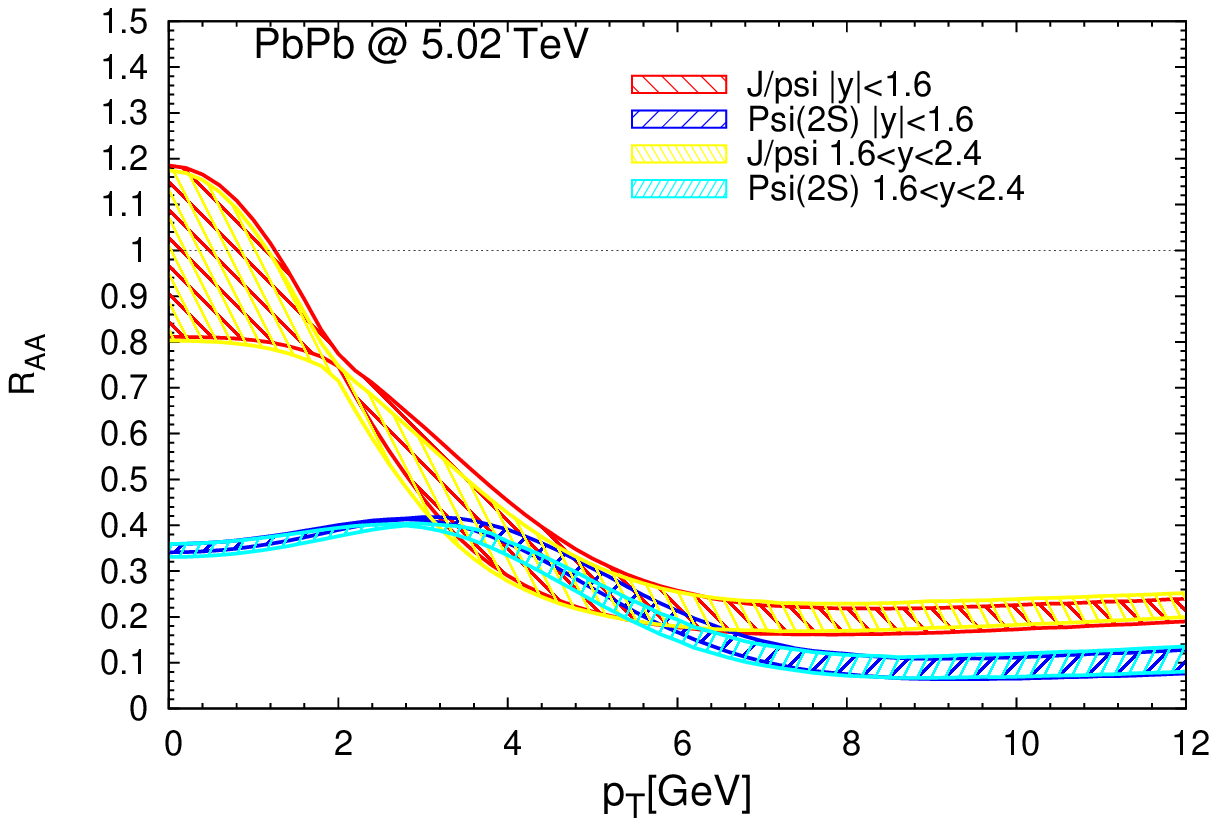}
\caption{Same as Fig.~\ref{fig_276} but for Pb-Pb collisions at $\sqrt{s_{NN}}$=5.02\,TeV.}
\label{fig_502}
\end{figure}
For our predictions at $\sqrt{s_{NN}}$=5.02\,TeV, we implement the following changes.
The charged particle multiplicity, {\it i.e.}, $S_{\rm tot}$, is increased by 22.5\%~\cite{Niemi:2015voa},
yielding an increase of $\sim$7\% for the initial temperature.
Using experimental~\cite{Andronic:2015wma} and theoretical~\cite{Cacciari:2012ny} results as guidance,
the charm cross section in pp, $\frac{\mathrm{d}\sigma_{c\bar{c}}}{\mathrm{d}y}$, is increased by
$\sim$40\%, from 0.65(0.59)\,mb to 0.92(0.84)\,mb at mid-(forward) rapidity, and likewise for charmonia.
Their initial $p_T$ spectra in pp are updated (somewhat harder than at 2.76\,TeV), and a 10\% additional shadowing is assumed. Finally, for the $J/\psi$
blast-wave spectra, we found that a somewhat lower temperature range of $T$=179-181\,MeV
(representing a time window around the mixed phase) better describes the preliminary ALICE
dimuon data for the $J/\psi$ $R_{AA}(p_T)$ at 5.02\,TeV as presented at this
meeting~\cite{Adam:2016rdg}.

The resulting $R_{\rm AA}$ double ratios for the different $p_T$ cuts show a trend of moving
closer together (see left panel of Fig.~\ref{fig_502}). Due to the increase in transverse flow,
more regenerated $J/\psi$ are pushed beyond the $p_T$$>$3\,GeV thus suppressing the red band,
while more $\psi(2S)$ are pushed beyond $p_T$$>$6.5\,GeV thus enhancing the blue band,
cf.~also the individual $R_{\rm AA}(p_T)$'s in the right panel of Fig.~\ref{fig_502}.

\section{Conclusions}
\label{sec_concl}
In summary, the application of a kinetic rate equation approach to charmonium
production in HICs suggests a scenario where $\psi(2S)$ states are regenerated
significantly later in the fireball evolution than $J/\psi$ mesons. This emerges
from inelastic reaction rates which are significantly larger for the $\psi(2S)$
than the $J/\psi$ in the later (hadronic) stages of the fireball. This
``sequential regeneration" is a direct consequence of sequential suppression
plus detailed balance. We have found that the phenomenological implications of
this scenario can help to explain the puzzling observation of the CMS
$\psi(2S)$-to-$J/\psi$ $R_{\rm AA}$ double ratio.
\\

{\ack This work is supported by the US National Science Foundation under grant no. PHY-1614484.}\\

\end{document}